\documentclass{article}
\usepackage{graphicx} 

\usepackage[table, x11names]{xcolor}
\usepackage{array, booktabs, boldline} 

\PassOptionsToPackage{table}{xcolor}
\usepackage{jheppub} 

\usepackage{caption}
\usepackage{subcaption}
\usepackage[all]{xy}
\usepackage[percent]{overpic}
\usepackage{slashed,physics}
\usepackage{wrapfig}
\usepackage{tabu}
\usepackage{diagbox}
\usepackage{mathrsfs,amsmath,amssymb,amsthm,amsfonts,tikz,graphicx,accents,hyperref, color,bbold}
\usepackage{dsfont,epiolmec, latexsym, stmaryrd, comment}
\usepackage{slashed,ccaption}
\usepackage{mathrsfs, calligra}
\usepackage{leftidx}
\usepackage{import}
\usepackage{multirow}
\usepackage{amsfonts}
\usepackage{pifont}
\usepackage{tabularx}
\usepackage[utf8]{inputenc}
\usetikzlibrary{intersections,calc}
\usepackage{tikz-3dplot}
\usepackage{ifthen}
\usetikzlibrary{arrows}
\usepackage{xcolor}

\usepackage{caption}

\newcommand{\AB}[1]{#1} 
\newcommand{\GJ}[1]{#1} 
\newcommand{\JM}[1]{#1} 
\usepackage{array}
\usepackage[percent]{overpic}
\usepackage{wrapfig}
\usepackage{bbm}
\usepackage{tabu}
\usepackage{slashed}
\usepackage{fancyhdr} 
\usepackage{amsmath}
\usepackage{amsfonts}
\usepackage{amssymb}
\usepackage{diagbox}

\hypersetup{ linktoc=all,
    colorlinks, linkcolor={blue}, 
    citecolor={red}, urlcolor={blue}
}

\graphicspath{{Images/}}

\definecolor{light gray}{RGB}{220,220,220}
\definecolor{dark purple}{RGB}{108,0,217}
\definecolor{pink}{RGB}{190,20,100}
\definecolor{orang}{RGB}{193,63,0}
\definecolor{green}{RGB}{11,98,17}
\definecolor{darkpink}{RGB}{153,0,76}
\definecolor{bluegreen}{RGB}{0,102,102}
\definecolor{greenlagan}{RGB}{0,102,0}
\definecolor{redgreen}{RGB}{102,102,0}
\definecolor{Redgreen}{RGB}{153,76,0}
\definecolor{vividviolet}{rgb}{0.62, 0.0, 1.0}
\definecolor{amaranth}{rgb}{0.9, 0.17, 0.31}
\definecolor{palatinateblue}{rgb}{0.15, 0.23, 0.89}
\definecolor{brightpink}{rgb}{1.0, 0.0, 0.5}
\definecolor{cornflowerblue}{rgb}{0.39, 0.58, 0.93}
\definecolor{deepcarminepink}{rgb}{0.94, 0.19, 0.22}
\definecolor{radicalred}{rgb}{1.0, 0.21, 0.37}
\usepackage{graphicx}
\usepackage{tikz}
\usetikzlibrary{arrows,chains,shapes,matrix,positioning,scopes}
\usetikzlibrary{decorations.markings}
\tikzstyle arrowstyle=[scale=1]
\tikzstyle directed=[postaction={decorate,decoration={markings,
    mark=at position .65 with {\arrow[arrowstyle]{stealth}}}}]
\tikzstyle reverse directed=[postaction={decorate,decoration={markings,
    mark=at position .65 with {\arrowreversed[arrowstyle]{stealth};}}}]
\usepackage{import}
\usepackage{accents}
\usepackage{mathrsfs,amsmath,amssymb,slashed}
\usepackage{multirow,multicol}
\usepackage{enumitem}
\usepackage[percent]{overpic}
\usepackage{slashed}

\usepackage{wrapfig}
\usepackage{tabu}
\usepackage{diagbox}
\usepackage{comment}
\usepackage{tikz}
\usepackage{mathrsfs,amsmath,amssymb,amsthm,amsfonts,graphicx,accents,hyperref,color}
\usepackage{leftidx}
\usepackage{import}
\usetikzlibrary{decorations.pathmorphing}
\DeclareFontFamily{OT1}{rsfs}{}

\DeclareFontShape{OT1}{rsfs}{m}{n}{ <-7> rsfs5 <7-10> rsfs7 <10->rsfs10}{} 

\DeclareMathAlphabet{\mycal}{OT1}{rsfs}{m}{n}
\title{Krylov Complexity and Spectral Form Factor for Noisy Random Matrix Models}
\author[a]{Arpan Bhattacharyya,}
\author[b,c,d]{S. Shajidul Haque,}
\author[b,e]{Ghadir Jafari,}
\author[b,d]{Jeff Murugan}
\author[b] {and Dimakatso Rapotu
}
\date{May 2023}

\affiliation{\it $^{a}$ Indian Institute of Technology, Gandhinagar, Gujarat 382355, India}
\affiliation{\it $^b$ Laboratory for Quantum Gravity \& Strings, Department of Mathematics and Applied Mathematics,
	University of Cape Town, South Africa
}
\affiliation{\it $^{c}$ Brac University, Dhaka, Bangladesh}

\affiliation{\it $^{d}$ National Institute for Theoretical and Computational Sciences (NITheCS), Private Bag X1, Matieland, South Africa}

\affiliation{\it $^e$ Department of Physics Education, Farhangian University, P.O. Box 14665-889, Tehran, Iran}
\emailAdd{abhattacharyya@iitgn.ac.in}
\emailAdd{shajid.haque@uct.ac.za }
\emailAdd{ghadir.jafari@uct.ac.za}
\emailAdd{Jeff.murugan@uct.ac.za}
\emailAdd{rptdim002@myuct.ac.za}

\abstract{We study the spectral properties of two classes of random matrix models: non-Gaussian RMT with quartic and sextic potentials, and RMT with Gaussian noise. We compute and analyze the quantum Krylov complexity and the spectral form factor for both of these models. We find that both models show suppression of the spectral form factor at short times due to decoherence effects, but they differ in their long-time behavior. In particular, we show that the Krylov complexity for the non-Gaussian RMT and RMT with noise deviates from that of a Gaussian RMT. We discuss the implications and limitations of our results for quantum chaos and quantum information in open quantum systems. Our study reveals the distinct sensitivities of the spectral form factor and complexity to non-Gaussianity and noise, which contribute to the observed differences in the different time domains.}

\begin{document}
\maketitle
\section{Introduction}

Nothing in Nature is isolated\footnote{Not even the Universe itself if Multiverse proponents are to be believed.}. Closed quantum systems are an idealization that, while teaching us a great deal about the principles of quantum mechanics, must eventually give way to {\it open quantum systems} that encode the interaction between quantum systems of interest and the environments that they find themselves embedded in \cite{2007LNP...717.....A,doi:10.1142/12402,Breuer:2002pc,2012oqs..book.....R,2019arXiv190200967L}. Unfortunately, since such systems are no longer unitary, conventional Hamiltonian methods to study their dynamics fail or need to be modified at the very least. In particular, the von Neumann equation,
\begin{equation*}
    \dot{\rho} = -\frac{i}{\hbar}[H,\rho]\,,
\end{equation*}
that describes the evolution of the density matrix of a closed quantum system is replaced by the Gorini–Kossakowski–Sudarshan–Lindblad (GKSL) (or, more commonly, the {\it Lindblad}) master equation \cite{Gorini:1975nb,1976CMaPh..48..119L},
\begin{equation*}
    \dot{\rho} = -\frac{i}{\hbar}[H,\rho] + \sum_{i}\gamma_{i}\left(L_{i}\,\rho \,L_{i}^{\dagger} - \frac{1}{2}\left\{L_{i}^{\dagger}L_{i},\rho\right\}\right)\,,
\end{equation*}
in which the system-environment interactions are encoded in the set of {\it jump operators}, $L_{i}$ and the non-negative coefficients $\gamma_{i}$ are the damping rates that control the strength of the environmental interaction. Open quantum systems generically thermalize by leaking information/energy into the ambient bath \cite{2015PhRvL.115v0401D}, and thermal systems, in turn, exhibit properties of a random matrix ensemble; unitary, orthogonal or symplectic, depending on the starting Hamiltonian \cite{alma99207451986006446}. The properties of such ensembles are captured by random matrix theory (RMT) \cite{alma99207451986006446}.\\  

\noindent
Since its introduction to physics by Wigner \cite{c5bd8f0f-2576-3f83-a184-791e55682183,90dbe650-d74b-3174-ae84-1c16c4e3fd5c,c9eb8278-b5e2-37bc-a322-8e81785f98ed}, RMT has proven to be a powerful tool in the study of statistical properties, such as energy spectra and level spacing distributions, of complex quantum systems \cite{1626a8bf-8479-36bf-b395-22e4def81091,10.1214/aoms/1177728846}. It has also been successfully applied across various subdisciplines, including nuclear and condensed matter physics, quantum chaos, and quantum information \cite{Akemann:2011csh,2016JMP....57a5215C,DAlessio:2015qtq,DiFrancesco:1993cyw,1997RvMP...69..731B,Guhr:1997ve}. However, most RMT models assume that the matrix elements are independent and identically distributed random variables with a Gaussian distribution \cite{Dyson:1972tm,alma99207451986006446}. This assumption may only be valid for some realistic quantum systems, especially those that are open to the environment and subject to noise and dissipation.\\

\noindent
In this article, we investigate two classes of modified RMT models that can potentially capture some aspects of open quantum systems: {\it non-Gaussian} RMT \cite{alma99207451986006446,Gaikwad_2019} and RMT with {\it noise} \cite{Cornelius_2022}. The non-Gaussian RMT models considered here are defined by perturbing the standard quartic potential in the matrix model with a sextic term\footnote{\GJ{In fact, in the Random Matrix Theory (RMT) literature, two distinct aspects are referred to as ``Gaussianity." The first pertains to the ensemble from which we select the matrices, while the second relates to the measure. In this context, we have chosen matrices from the conventional Gaussian ensemble. However, the measure (or potential) we employ, which includes quartic or sextic terms, is also termed ``non-Gaussian."}}.  \AB{It has richer phase structure than the Gaussian one \cite{alma99207451986006446,Gaikwad_2019}. On the other hand, random matrix models with noise are defined by adding a random perturbation to a Gaussian RMT model, which can mimic the effect of environmental noise on the system. In particular, motivated by the studies done in \cite{Cornelius_2022} we will be focusing a particular model where the noise in the energy spectrum can be generated  by the fluctuations of the system's Hamiltonian in time. These kind of energy diffusion processes characterized by energy dephasing arise in plethora of interesting physical scenario e.g. in random quantum measurements, through the errors of the clock used for timing the
evolution of a quantum system. Interested readers are referred to \cite{PhysRevLett.52.1657,PhysRevA.96.032124,PhysRevA.59.3236,PhysRevA.44.5401,PhysRevD.67.025007,PhysRevLett.118.140403,Cornelius_2022} for more details.} We compare and contrast these two classes of models and analyze their spectral properties to extend our understanding of the interplay between dissipation and chaos in quantum systems.\\

\noindent
Two of the main tools that we will use in our analysis are quantum complexity \cite{Parker_2019} and the spectral form factor \cite{1997PhRvE..55.4067B} associated with the system's Hamiltonian. Quantum complexity - a notion borrowed from computer science - measures the difficulty of preparing or simulating a quantum state using a quantum computer. There are, by now, several complementary notions of quantum complexity, depending on the choice of the computational model and the resources considered. One notion of complexity that has recently been developed is the {\it Nielsen complexity} \cite{NL1,NL2,NL3}, defined as the minimum number of elementary gates needed to approximate a given unitary transformation with a certain accuracy \footnote{Interested readers are referred to some of the works done in the context of Nielsen complexity for quantum field theory and quantum many-body systems \cite{Jefferson,Chapman:2017rqy,Bhattacharyya:2018wym,Caputa:2017yrh,me1,Bhattacharyya:2018bbv,Hackl:2018ptj,Khan:2018rzm,Camargo:2018eof,Ali:2018aon,Caputa:2018kdj,Guo:2018kzl,Bhattacharyya:2019kvj,Flory:2020eot,Erdmenger:2020sup,Ali:2019zcj,Bhattacharyya:2019txx,cosmology1,cosmology2, Caceres:2019pgf,Bhattacharyya:2020art,Liu_2020,Susskind:2020gnl,Chen:2020nlj,Czech:2017ryf,Chapman:2018hou,Geng:2019yxo,Guo:2020dsi,Couch:2021wsm,Erdmenger:2021wzc,Chagnet:2021uvi,Koch:2021tvp,Bhattacharyya:2022ren,Bhattacharyya:2023sjr,Bhattacharyya:2022rhm,Bhattacharyya:2021fii,Bhattacharyya:2020iic,ShashiPRB,casagrande2023complexity,Craps:2023rur,Haque:2021hyw,Haque:2021kdm}. This list is by no means exhaustive. Interested readers are referred to these reviews \cite{Chapman:2021jbh, Bhattacharyya:2021cwf}, and references therein for more details.}. Another notion of quantum complexity, and one of the foci of this article, is {\it Krylov complexity} \cite{Parker_2019, Balasubramanian:2022tpr}, associated with the growth of an operator, or spread of a state under time-evolution in a certain (sub)space spanned by orthonormal basis vectors known as `Krylov space'. A recent surge of work in this area has demonstrated that these notions of quantum complexity capture some essential features of quantum chaos and quantum ergodicity. Interested readers are referred to some of these references, which are by no means exhaustive \cite{Parker:2018yvk,Barbon:2019wsy,Avdoshkin:2019trj,Cao:2020zls,Jian:2020qpp,DymarskyPRB2020,PhysRevLett.124.206803,Yates2020,Rabinovici:2020ryf,Rabinovici:2021qqt,Yates:2021lrt,Yates:2021asz,Dymarsky:2021bjq,Noh2021,Trigueros:2021rwj,https://doi.org/10.48550/arxiv.2207.13603,Fan_2022,Kar_2022,https://doi.org/10.48550/arxiv.2109.03824,PhysRevE.106.014152,https://doi.org/10.48550/arxiv.2204.02250,Bhattacharjee_2022,Bhattacharjee_2022a,Du:2022ocp, Banerjee:2022ime,
https://doi.org/10.48550/arxiv.2205.12815,H_rnedal_2022,https://doi.org/10.48550/arxiv.2208.13362,Rabinovici:2022beu,Alishahiha:2022anw, Avdoshkin:2022xuw,Camargo:2022rnt,Kundu:2023hbk,Rabinovici:2023yex,Zhang:2023wtr,Nizami:2023dkf,Hashimoto:2023swv,Nandy:2023brt,Balasubramanian:2022tpr, Caputa2022PRB, Caputa:2022yju, https://doi.org/10.48550/arxiv.2208.10520,Balasubramanian:2022dnj,Erdmenger:2023shk,Bhattacharya:2022gbz,Bhattacharjee:2022lzy,Bhattacharya:2023zqt,Liu:2022god,Chattopadhyay:2023fob,Pal:2023yik,Patramanis:2023cwz, Bhattacharyya:2023dhp, Caputa:2023vyr,Camargo:2023eev,Iizuka:2023pov,Vasli:2023syq}. We will show here that Krylov complexity for non-Gaussian RMT and RMT with noise is different from that of Gaussian RMT, and we provide a physical interpretation of this difference. To the best of our knowledge, this is the first time that Krylov complexity is computed for such \textit{RMT models with noise} \footnote{Please refer to \cite{Balasubramanian:2022dnj, Erdmenger:2023shk} for recent studies of Krylov complexity for certain RMT without any noise.}.\\

\noindent
One of the defining characteristics of random matrices is their {\it spectral rigidity}, the tendency of the eigenvalues of a \JM{random matrix  to spread out, resulting in a particular statistical distribution of eigenvalues \cite{alma99207451986006446,Garcia-Garcia:2016mno}. However, the degree of this rigidity is not universal, but depends on the symmetry and the distribution of the matrix entries. For example, Gaussian random matrices, such as the Gaussian orthogonal ensemble (GOE), Gaussian unitary ensemble (GUE), and Gaussian symplectic ensemble (GSE), have a high degree of spectral rigidity, meaning that their eigenvalue spacing distribution deviates significantly from the Poisson distribution. On the other hand, some other random matrix ensembles, such as the Ginibre ensemble, have a low degree of spectral rigidity, meaning that their eigenvalue spacing distribution is close to the Poisson distribution. This is due essentially to the fact that matrices from these ensembles lack symmetry constraints and their entries are independent and identically distributed. One well-known measure of spectral rigidity is the Dyson-Mehta $\Delta_{3}$ statistic, defined as the minimum mean-square deviation of the integrated eigenvalue density from a linear function over an interval of length $L$ of the unfolded spectrum \cite{10.1063/1.1704008}. For example, for a regular system with Poisson statistics,  and no correlation between levels, the ensemble-averaged Dyson-Mehta statistic grows linearly with $L$ as $\overline{\Delta_{3}} = L/15$, whereas Gaussian ensembles are more rigid and the growth is logarithmic,
$\overline{\Delta_{3}} = \frac{1}{\pi^{2}}\left[\log(2\pi L)+\gamma - \frac{5}{4}\right] - \frac{1}{8} + \mathcal{O}(L^{-1})$\,, for large $L$ and for a GOE.} \\

\noindent
The second goal of this paper is to compute and analyze spectral features of non-Gaussian random matrices and noisy random matrix models. Toward this end, we compute the spectral form factor \cite{1997PhRvE..55.4067B} - defined as (the Fourier transform of) the two-point correlation function of eigenvalues of the Hamiltonian - that encodes the long-range spectral fluctuations of the quantum system. It is closely related to the level spacing distribution and the level number variance \cite{Papadodimas:2015xma, Cotler:2016fpe, Garcia-Garcia:2016mno}. It is essentially a measure of the correlation between energy eigenvalues of a Hamiltonian system and can be used to characterize the quantum chaos of the system by comparing it with the Random Matrix Theory (RMT) predictions \cite{Cotler:2016fpe, Garcia-Garcia:2016mno,Dyer:2016pou,Krishnan:2016bvg,Balasubramanian:2016ids,delCampo:2017bzr}. The spectral form factor usually shows a dip at a certain time scale, followed by a ramp and a plateau. The dip is caused by the disconnected pair correlations between density of states, while the ramp and plateau are caused by the connected pair correlations between density of states. We will show that the spectral form factors for non-Gaussian RMT \cite{Gaikwad_2019} and RMT with noise \cite{Cornelius_2022} exhibit  behaviour different from that of Gaussian RMT, and we will explain how these behaviours reflect the non-Gaussianity and noise in the models.\\

\noindent
The paper is organized as follows: In Section 2, we introduce the definitions and properties of the perturbed Gaussian random matrix models that are the focus of this article. In Section 3, we compute and analyze the spectral form factors for these models using a combination of analytical and numerical methods. Section 4 is taken up with the computation of the Lanczos coefficients and the associated Krylov complexity for these models. In Section 5, we summarize our main results and discuss their implications and limitations.\\

\section{Our Models} \label{sec2}
In this section, we provide a brief overview of the models employed in the subsequent sections of the paper. As previously mentioned, one of the canonical methods for characterizing quantum chaos involves comparing the spacing between nearest-neighbour energy eigenvalues with that of an ensemble of random matrices (RMT). Consequently, determining the spectral properties of various random matrix models is of great importance for a broader understanding of quantum chaos. A key such property of these models is that they possess a characteristic time scale above which universal correlations between energy levels emerge \cite{1986JETP...64..127A, 2015CMaPh.333.1365E,Garcia-Garcia:2018ruf,Gharibyan:2018jrp}. This realization led to the development of probes for characterizing late-time chaos, such as spectral form factors. The spectral form factor displays universal behaviour for Gaussian RMT and has been extended to non-Gaussian Models in \cite{Gaikwad_2019}.\\

\noindent
One of the primary objectives of this article is to investigate the universality of this chaotic behaviour and compare it with the features of Krylov complexity. We will initially examine and compare the behaviour of the spectral form factor and Krylov complexity for both non-Gaussian RMT and RMT models with noise. Below, we provide a brief description of both systems.

\subsection*{Non-Gaussian Random Matrix Models}
We will start with the simplest non-Gaussian models following  \cite{Gaikwad_2019}. The model is characterised by a quartic and sextic potential, 
\begin{equation} \label{nongauss}
	V(M)=\frac{1}{2} M^2+\frac{g}{N} M^4+\frac{h}{N^2} M^6
\end{equation}
Here we will take the matrices $M$ from a \textit{Gaussian Unitary ensemble} (GUE) \footnote{\GJ{As the matrices are drawn from the Gaussian Unitary Ensemble (GUE), we can apply the standard tools to generate this ensemble. Specifically, in our calculations, we have utilized the \texttt{GaussianUnitaryMatrixDistribution} function from \textbf{Mathematica}.}}. Following \cite{1993lneq.book..567B,Eynard:2015aea}, it can be shown that this model possesses a line of critical points in the $(g,h)$ plane, given parametrically by 
\begin{equation*}
    g=\frac{3-2 b^2}{12 b^4}\,\quad\mathrm{and}\quad h=\frac{b^2-2}{60 b^6}\,,
\end{equation*}
where the parameter $b$ satisfies the sixth order polynomial $ 60\,h\,b^6 +12\,g\,b^4+b^2-1 =0$.
After writing $g = g(h)$ on this critical line one can show that, for a fixed value of $h$, when $g > g_{c}(h)$, there are one, two and three cut solutions. However, for $g<g_{c}(h)$ and a fixed value of $h$, there are only two and three-cut solutions. When $b^2=3,$ there is tri-critical point where $g=-\frac{1}{36}, h=\frac{1}{1620}.$ For the rest of this paper, we will consider values of $g$ and $h$ above this tri-critical value. 

\subsection*{Random Matrix Models with Noise}

Next, we turn our attention to another random matrix model in the context of an open quantum system. While the addition of the sextic term to the matrix model probes its chaoticity away from Gaussianity, we would to extend this and test the robustness of the quantum chaotic nature of the underlying theory in the presence of decoherence \cite{delCampo:2019qdx,Cornelius_2022,Zhou:2023qmk}. In particular, we would like to test the reported observation that, under certain circumstances, the presence of decoherence in a system enhances the underlying quantum chaotic behaviour \cite{Xu:2018bhd}. Specifically, we ask whether the dip-ramp-plateau behaviour of the spectral form factor that characterises chaotic behaviour is modified at all in the presence of decoherence. As discussed briefly in the introduction, for this case, time evolution is  non-Hermitian and governed by the Lindbladian through the evolution equation, \cite{Gorini:1975nb,1976CMaPh..48..119L},
\begin{align} \label{lindblad}
\dot{\rho}=-i\,[H_0,\rho]+\sum_i\,\gamma_i\,\Big(L_i\,\rho\,L_i^{\dagger}-\frac{1}{2}\{L_i^{\dagger}L_i,\rho\}\Big)\,,
    \end{align}
where $\rho$ denotes the density matrix of the open system, and the overdot denotes its time derivative. For the purposes of our study, the system Hamiltonian $H_{0}$ will be identified with the matrix model potential $V(M)$. The $\gamma_i \geq 0$ are decay rates for the system and the $L_i's$ are so-called \textit{jump operators}. At this point, these can be arbitrary functions of $H_0.$ Up to a normalization, we can write them as  $L_i=L_{i}^{\dagger}\propto w(H_0)$ where $w(H_0)$ is an arbitrary complex function of $H_{0}.$ Motivated by studies carried out in \cite{Cornelius_2022}, we will mainly focus on the scenario where the underlying physical process is dictated by energy dephasing \cite{PhysRevLett.52.1657,PhysRevA.96.032124,PhysRevA.59.3236,PhysRevA.44.5401,PhysRevD.67.025007,PhysRevLett.118.140403,Cornelius_2022}, and compare the behaviour of the spread complexity with that of the spectral form factor. The latter is known to remain a useful probe of quantum chaos even for open systems in the presence of energy dephasing \cite{Cornelius_2022}. For this case study, we will choose $w(H_0)=H_0^{\delta}$ where $\delta$ can be an integer or a half-integer \cite{Cornelius_2022}. This choice introduces Gaussian noise in the spectral form factor \cite{Cornelius_2022}. For most subsequent studies, we will choose $\delta=1$, resulting in a sub-Gaussian filter function in the spectral form factor. 

\section{Spectral Form Factor}
To analyze the spectral properties of our models discussed in the previous section, we begin by computing the spectral form factor (SFF). Originally introduced as a probe for analyzing the spectrum of quantum systems, it has developed into an indispensable probe of quantum chaotic systems \cite{Cotler:2016fpe}. It is defined in terms of the analytically continued partition function as,
\begin{equation} \label{SFFdefinition}
\left|\frac{Z(\beta+i\,t)}{Z(\beta)}\right|^2=\sum_{m, n} \frac{e^{-\beta\left(E_m+E_n\right)} e^{-i t\left(E_m-E_n\right)}}{Z^2(\beta)}
\end{equation}
\noindent
As such, the SFF interpolates between early-time probes such as the out-of-time order correlator (OTOC) and more standard RMT measures such as the $r$-statistics of the spectrum. This makes it particularly well-suited for the study of systems transitioning between integrable and chaotic behaviour \cite{Cotler:2016fpe}. However, except for some special cases such as single-particle bosonic quantum mechanics where it can be shown that the two-point SFF is obtained by averaging the four-point OTOC over the Heisenberg group \cite{deMelloKoch:2019rxr}, computing the SFF is a difficult task, compounded by various subtleties inherent to the spectral analysis of the chaotic Hamiltonian. Unfortunately, neither of our models fall allow for an analytic computation of the SFF and so must be treated numerically.

\subsection{SFF for Non-Gaussian Models}
\begin{figure}[htb!]
	\centering
\includegraphics[scale=1]{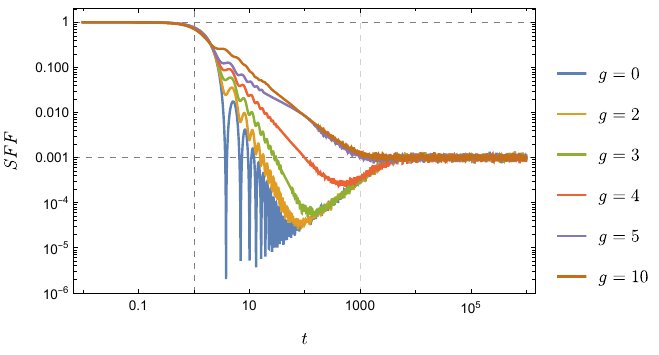}
\caption{SFF for quartic potential for various values of the coupling.}
\label{fig:plt1avg}
\end{figure}

In this section, we will study the SFF for non-Gaussian random matrix systems mentioned in (\ref{nongauss}). This computation starts with the diagonalization of the Hamiltonian for the energy eigenvalues. These are then substituted into (\ref{SFFdefinition}) and the SFF computed over an ensemble of realizations of the Hamiltonian since the SFF is not particularly well-behaved for a single realization.  For example, and to benchmark our numerical methods, the SFF computed for just the quartic term turned on in (\ref{nongauss}) is shown in Fig.~(\ref{fig:plt1avg}).
This plot is generated for a matrix model with $N\times N$ matrices with $N=1000$ and averaged over $n=100$ realizations. To understand the role of two non-Gaussian parameters ($g, h$) in (\ref{nongauss}), we first turn off the sextic parameter $h$ and compute the  SFF over a range of the quartic coupling $g$. Observe that adding quartic couplings does not change the early-time behaviour, the saturation magnitude at a late time, nor the saturation time scale of the Gaussian model. However, as we increase the coupling $g$, the dip time increases relative to the Gaussian model, as evident from Fig.~(\ref{fig:plt1avg}). There is a value of the coupling $g_c\approx 6$ at which the dip and the subsequent ramp disappear. Note also that including the quartic non-Gaussian term does not change the slope of the ramp 
but does change its length.\\

\begin{figure}[t!]
    \centering
    \includegraphics[scale=1]{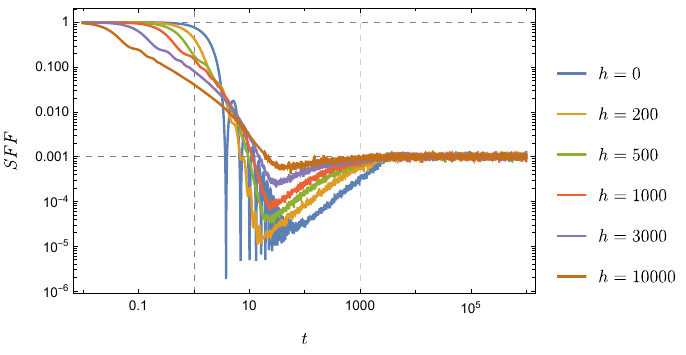}
    \caption{SFF for Sextic potential for various values of the coupling.}
    \label{fig:pltavg1}
\end{figure}

\noindent
Next, we turn off the quartic coupling and turn on the sextic term with coupling $h$. As in the quartic case, the coupling $h$ does not change the Gaussian model's saturation magnitude or time. It does, however, change the early-time behaviour. Another notable difference from the quartic case is that even though the dip time does not change appreciably with $h$, the slope of the ramp and the magnitude of SFF at the dip change considerably with increasing $h$. However, similar to the quartic case, there is a value of $h_c$ at which the dip and subsequent ramp disappear. This is demonstrated in the Fig.~(\ref{fig:pltavg1}).\\

\newpage

\begin{figure}[htb!]
    \centering
\includegraphics[scale=1]{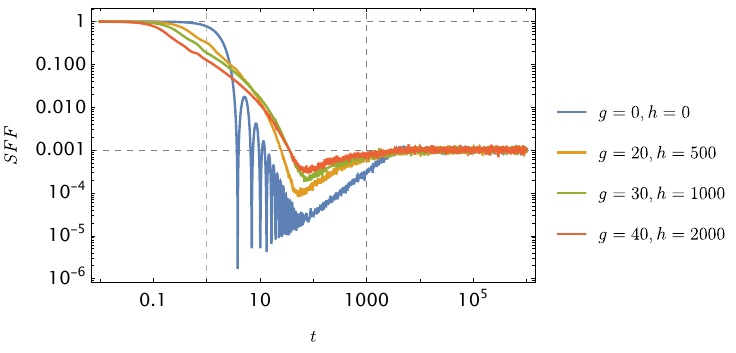}
    \caption{SFF for Non-Gaussian RMT with both quartic and sextic potential for different values of parameters.}
    \label{fig:pltghavg1}
\end{figure}

\noindent
Finally, turning on both couplings $(g, h)$ we compute the SFF for the entire model for different values of $g$ and $h$ and plot the result in Fig.~(\ref{fig:pltghavg1}). Interestingly for this extended model, the magnitude of the SFF at the dip may be tuned with relative ease up to its saturation level by  appropriate choice of the parameters $g$ and $h$. 

\subsection{Random Matrix Model with Noise}

How does the behaviour of the SFF change if, instead, we introduce noise into a quantum system such as the the random matrix model with Gaussian noise, introduced in Sec.~(\ref{sec2})? To be concrete, we will focus mainly on the physical process dictated by energy dephasing and choose the jump operators to be proportional to $H_0$. Then making use of a  Hubbard–Stratonovich transformation, the SFF in the presence of the noise has the integral representation \cite{Xu_2021},
\begin{equation}
	F_t=\frac{1}{2 \sqrt{\pi \gamma t}} \int_{-\infty}^{+\infty} d \tau e^{-\left(\frac{\tau-2 t}{2 \sqrt{\gamma t}}\right)^2} g_\beta(\tau),
\end{equation}
where $g_{\beta}$ is the usual SFF for the model without the noise, 
$$
g_{\beta} (\tau)=\left|\frac{Z(\beta+i \tau)}{Z(\beta)}\right|^2\,,
$$
and $\gamma$ is the strength of the noise as defined in (\ref{lindblad}). It is also related to the decay rate. 
The effect of noise on the SFF for the Gaussian model is exhibited in Fig.~(\ref{fig:pltgamma}) and can be summarised as follows:
\begin{figure}[htb!]
	\centering
\includegraphics[scale=1]{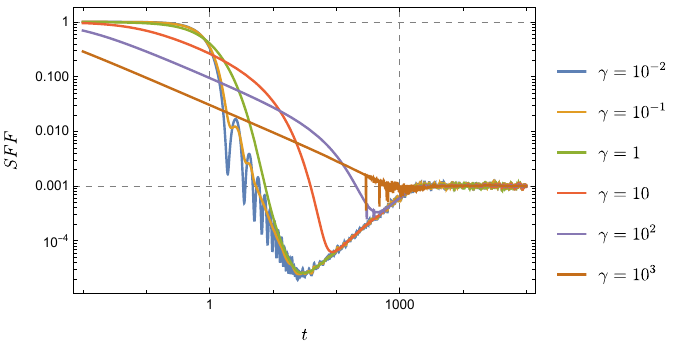}
	\caption{SFF for Gaussian random matrix model with Gaussian noise.}
	\label{fig:pltgamma}
\end{figure}
the presence of noise alters the early-time behaviour, but does not affect the saturation magnitude, saturation time, or slope of the ramp. It is important to note that while some aspects of the impact of the introduction of noise align with the non-Gaussian terms, not all of its effects can be replicated by these non-Gaussian terms alone. Consequently,  noise alone cannot generate an equivalent effect to non-Gaussianity (quartic and sextic) on its own, as one can see by comparing Fig.~(\ref{fig:pltgamma}) and Fig.~(\ref{fig:pltghavg1}).\\

\begin{figure}[htb!]
    \centering
    \includegraphics[scale=1]{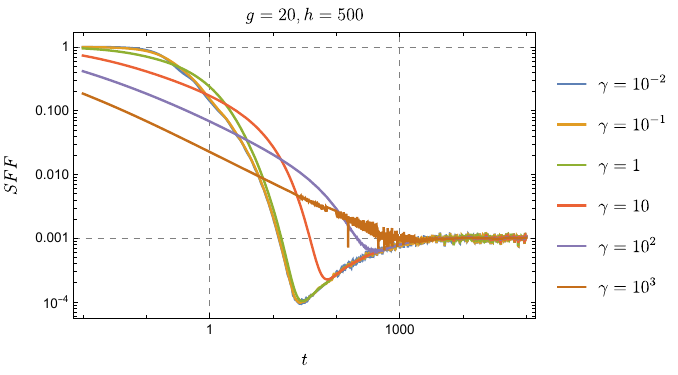}
    \caption{SFF for Non-Gaussian RMT with Gaussina noise.}
    \label{fig:pltghgamma1}
\end{figure}

\noindent
Before concluding this section, let's briefly compare the effect of non-Gaussianity on the SFF in the presence of noise. Here again, we numerically compute the spectral form factor and plot its time-evolution in Fig.~(\ref{fig:pltghgamma1}). For these systems, we observe that noise effectively reduces the length of the ramp. This is true for both the Gaussian and non-Gaussian random matrix models. To understand these observations, it is worth keeping in mind that physically, the dip in the SFF means that the system has less correlation than expected by RMT; conversely, the ramp and plateau mean that the system has more correlation than expected by RMT \cite{Mukherjee:2020sbt}. Consequently, the absence of a dip in the SFF could either mean that the system is too chaotic - if, for example, the system has a very large level spacing - or not chaotic enough - if, for example, the system has some symmetry or is integrable - to match the RMT predictions. Evidently, then in both classes of systems (non-Gaussian and noisy) increasing the couplings brings with it an increase in the chaoticity of the system. To test this hypothesis, we require some independent measure which, in our case is provided by quantum complexity.

\section{Krylov Complexity} \label{sec4}
Having explored the spectral properties of these two types of random matrix models using the SFF, we now turn our attention to another probe, a quantum information-theoretic quantity, called \textit{Krylov Complexity} \cite{Parker_2019}. Like its more familiar counterpart, Neilsen complexity, there are two notions of Krylov complexity. The first essentially measures operator growth in Hilbert space through the development of nested commutators with the system Hamiltonian. In this article, we focus the more recently developed notion of the complexity associated with the spread of a state in the Krylov basis under time evolution and called, appropriately enough, \textit{spread complexity} \cite{Balasubramanian:2022tpr}. Let's briefly recall the essential ingredients behind the computation of the spread complexity. A quantum mechanical state of any closed system evolves according to the Schrodinger equation, 
\begin{equation*}
    i\frac{d}{dt}|\psi(t)\rangle = H_{0}|\psi(t)\rangle\,,
\end{equation*}
so that the state of the at the time $t$ is given by the series,
$$
|\psi(t)\rangle=\sum_{n=0}^{\infty} \frac{(-i t)^n}{n !}\left|\psi_n\right\rangle,
$$ 
where the
$$
\left|\psi_n\right\rangle \equiv H_0^n|\psi(0)\rangle\,.
$$
In this sense, the $\left|\psi_n\right\rangle$ may be considered as a basis spanning the space of states. However, this set is neither orthogonal nor normalized in general. To remedy this, we can apply the usual Gram–Schmidt orthogonalization procedure\footnote{For more details about this {\it Lanczos algorithm} \cite{Lanczos:1950zz} to the $\left|\psi_n\right\rangle $, used to generate such an orthonormal basis, we refer the interested reader to \cite{Viswanath1994}.}. The result is an ordered, orthonormal basis,
$$
\mathcal{K}=\left\{\left|K_n\right\rangle: n=0,1,2, \cdots\right\}
$$
that spans the subspace of the Hilbert space explored by time development of 
$$
|\psi(0)\rangle \equiv\left|K_0\right\rangle\,.
$$
The basis 
$\mathcal{K}$ is conventionally known as the Krylov basis. Expressed in the Krylov basis any Hermitian matrix $H_0$ takes a tri-diagonal form, 
$$
H_t=\left(\begin{array}{cccccc}
a_0 & b_1 & & & & \\
b_1 & a_1 & b_2 & & & \\
& b_2 & a_2 & b_3 & & \\
& & \ddots & \ddots & \ddots & \\
& & & b_{N-2} & a_{N-2} & b_{N-1} \\
& & & & b_{N-1} & a_{N-1}
\end{array}\right)
$$
At this point, it is interesting to note that, as our initial state $|\psi(0)\rangle$ belongs to the Krylov basis, then the orthogonalization procedure can be solved if we can find a similarity transformation $O$ such that it brings $H_0$ to its tridiagonal form $H_t$ {\it i.e.}, 
\begin{equation} O\,H_0\,O^{T}=H_t\,, \label{similarity}
\end{equation}
with $O|\psi(0)\rangle = |\psi(0)\rangle$.
As long as $H_0$ is Hermitian, this similarity transformation always brings it to a tri-diagonal form \cite{garcia_horn_2017}. We will make use of this fact for our subsequent computations and can be implemented this via \texttt{Mathematica}'s inbuilt  \texttt{HessenbergDecomposition} module \cite{reference.wolfram_2022_hessenbergdecomposition}. \\

\noindent
Now, if we express $H_0$ in the Krylov basis and consider the initial state $|\psi(0)\rangle$ as one of the basis states \textit{i.e.}$\left|K_0\right\rangle=|\psi(0)\rangle$, then the action of the time-evolution operator $\exp(-iH_{0}t)$ on the initial state is simply, 
\begin{equation*}
    e^{-iH_{0}t}|K_{0}\rangle = \sum_{n=0}^{\infty} \psi_n(t) \left|K_n\right\rangle = |\psi(t)\rangle\,.
\end{equation*}
Finally then, the spread complexity is defined through,
\begin{equation} \label{eq:complexity}
\mathcal{C}(t)=\sum_n n\left|\psi_n(t)\right|^2=\langle \psi(t)|\widehat{K}|\psi(t)\rangle,\quad \widehat{K}=\sum_n\, n\, |K_n\rangle\langle K_n|
\end{equation}
It has been shown in \cite{Balasubramanian:2022tpr} that this quantity is minimized when the $|\psi(t)\rangle$ is expanded in terms of the Krylov basis relative to over any other possible basis. In the following sections, we will tri-diagonalize the random matrix Hamiltonians discussed in Sec.~(\ref{sec2}) and, using (\ref{eq:complexity}) compute the spread complexity as a function of time each of the two classes of matrix models.
\subsection{Krylov Complexity for Non-Gaussian RMT}
As with the SFF computation, we take as the system Hamiltonian
\begin{equation}
	H_{0}=\frac{1}{2} M^2+\frac{g}{N} M^4+\frac{h}{N^2} M^6\,.
\end{equation}
Following the procedure outlined above, the Krylov (spread) complexity is computed by
\begin{itemize}
    \item putting this Hamiltonian into its tri-diagonal form, $H_{t}$
    \item constructing the associated operator $U(t) = \exp(-i H_{t}t)$ and applying it to an initial state $\ket{\psi(0)} = \ket{K_0}$ to get the time-evolved state $\ket{\psi(t)}$ in the Krylov basis, 
    \item computing the complexity as the expectation value of the operator $\widehat{K}=\sum_n\, n\, |K_n\rangle\langle K_n|$ with respect to $\ket{\psi(t)}$, and
    \item averaging over an ensemble of different realizations of the system (characterized by fixed couplings and matrices $M$ drawn from an appropriate ensemble) to account for statistical variations.
\end{itemize}

\begin{figure}[htb!]
    \centering
    \includegraphics[scale=0.68]{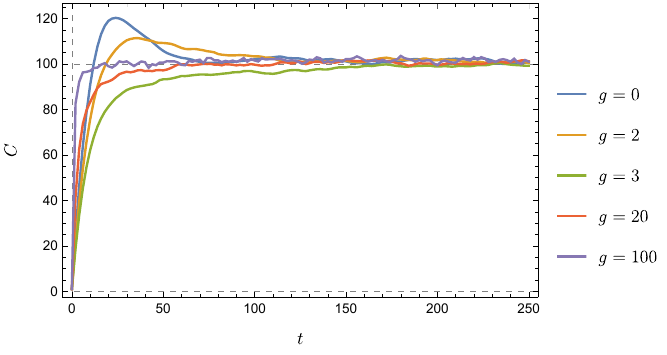}
     \includegraphics[scale=0.68]{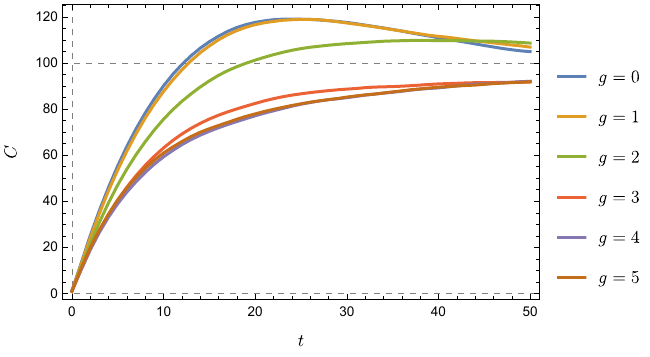}
    \caption{Evolution of Spread Complexity for quartic RMT. While the left panel shows the entire time evolution of the complexity, the right panel only shows the early-time behaviour. }
    \label{fig:complexity-g}
\end{figure}

\begin{figure}[b!]
    \centering
    \includegraphics[scale=.6]{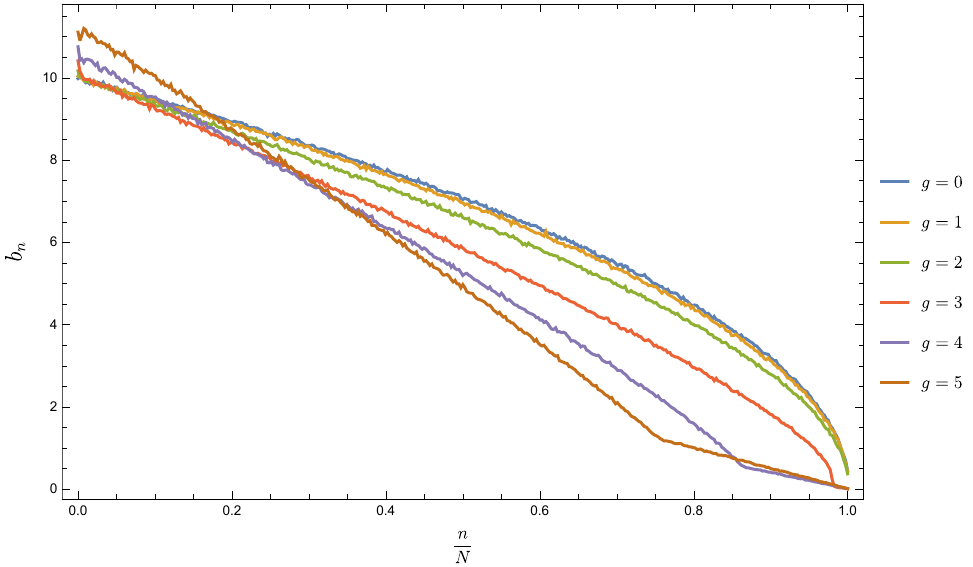}
    \caption{$b_n$ vs $n$ for different values of $g$}
    \label{fig:pltbng}
\end{figure}

\begin{figure}[b!]
    \centering
    \includegraphics[scale=0.82]{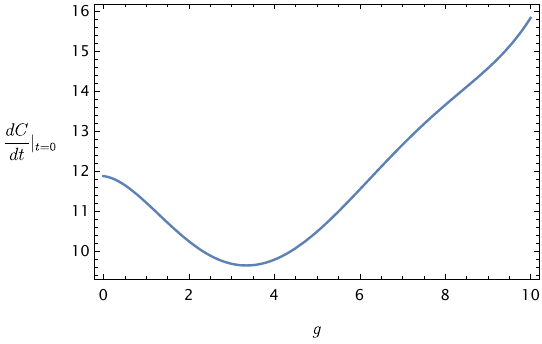}
     \includegraphics[scale=0.85]{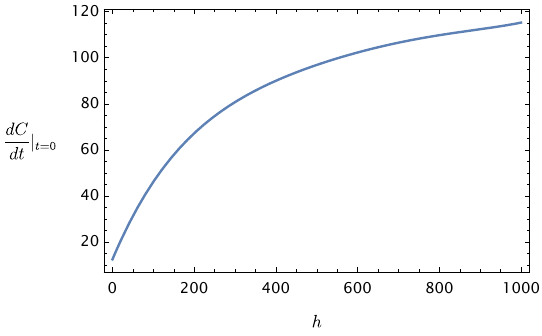}
    \caption{Rate of change of spread complexity at early time for quartic (left) and sextic (right) RMT.}
    \label{fig:crategh}
\end{figure}

\noindent
Let's first consider the effect of quartic coupling $g$ (setting $h=0$) on the complexity. The evolution of complexity for different values of the coupling $g$ is displayed in Fig.~(\ref{fig:complexity-g}). A few observations are worth pointing out. \begin{itemize}\item First; the saturation times and saturation magnitude of the complexities are approximately the same for different values of $g$, as evident from Fig.~(\ref{fig:complexity-g}). \item Second; we note that the Gaussian model ($g=0$) reaches a maximum before it decreases and saturates. This is consistent with\JM{ the observation made in \cite{Balasubramanian:2022tpr,Balasubramanian:2022dnj,Erdmenger:2023shk}} that (i) Krylov complexity exhibits a characteristic (and universal) peak and saturation structure in chaotic systems and (ii) the Gaussian RMT is chaotic. More curiously, this peak-saturation structure persists for $g=2$ but then dies out with increasing $g$ as evident from Fig.~(\ref{fig:complexity-g}). We cross-check this observation by computing the Lanczos $a_n$ and $b_n$ coefficients numerically. The results for the ensemble-averaged $b_{n}$ coefficients are displayed in Fig.(\ref{fig:pltbng}). Recalling that the statistics of the Lanczos $b_n$ coefficients, which are the off-diagonal elements of the tridiagonalized Hamiltonian, reveals some properties of the original operator. For example, if the original Hamiltonian is drawn from an invariant ensemble, then the $b_n$ coefficients are correlated and will follow a particular distribution depending on the ensemble as is the case above for $g=0$ where the ensemble-averaged $b_{n}\sim\sqrt{1-\frac{n}{N}}$. As the quartic coupling $g$ is increased, the distribution retains this basic profile until $g_{c}\approx 3$ when the profile changes qualitatively, developing a kink, that gets more pronounced as $g$ is increased beyond $g_{c}$. This agrees qualitatively with the critical value of the coupling in the spread complexity. \item Third; the linear slope that characterises the early-time behaviour of the Krylov complexity changes with the coupling. Specifically, the slope appears to decrease with increasing coupling until a minimum around $g_{c}\approx 3$ and then increases again. We verify this by plotting the slope of the complexity ($d\mathcal{C}/dt$) as a function of the quartic coupling in Fig.~(\ref{fig:crategh}). Here it is clear that around $g_{c}$,
\begin{equation}
\frac{d\mathcal{C}}{dt}\sim (g-g_c)^2\,.
\end{equation} 
\end{itemize}
\begin{figure}[htb!]
    \centering
\includegraphics[scale=0.80]{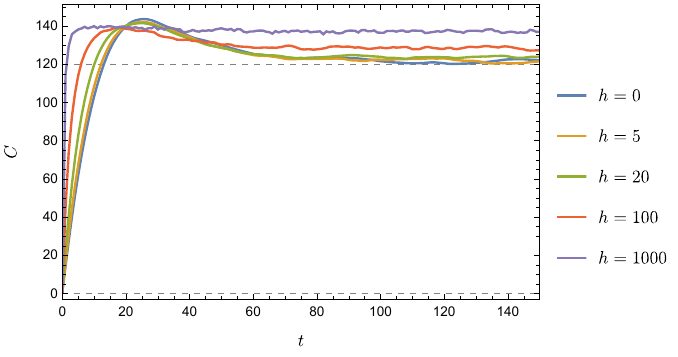}
    \caption{Spread Complexity for different values of sextic coupling $h\,.$}
    \label{fig:complexity-h}
\end{figure}
\noindent
This behaviour of the Krylov complexity for a quartic perturbation should be contrasted with the sextic perturbation. To this end we isolate the effects of the latter by turning off the quartic coupling $g$ and consider only the sextic coupling $h$. Fig.~(\ref{fig:complexity-h}) displays the complexity for different values of $h$. In this case, although the peak height decreases and saturation height increases with increasing $h$, the peak-saturation profile persists over a wide range of coupling strengths (up to approximately $h=1000$). We note also that, unlike the quartic case, the slope of the Krylov complexity is monotonic in the sextic coupling as shown in Fig.~(\ref{fig:crategh}).\\

\subsection{Krylov Complexity for RMT with Noise}
Now we extend our analysis of spread complexity for the RMT with Gaussian noise \footnote{We refer the reader to references \cite{Bhattacharya:2022gbz,Bhattacharya:2023zqt,Bhattacharjee:2022lzy,Liu:2022god} for studies of the Lancoz algorithm as well as its modifications and associated spread complexity for certain open quantum systems.}. Note that since this is an open quantum system, its time evolution is governed by the Lindblad equation \eqref{lindblad}. To set up the computation of the spread complexity, we first use the \textit{state-operator mapping} to associate the density operator with a state in a doubled Hilbert space \cite{1972RpMP....3..275J,CHOI1975285,PhysRevA.87.022310}. The time-evolution of the associated effective state is then generated by the operator
$e^{-i t\mathcal{L}_{o}}$, where $\mathcal{L}_o$ is the  Lindbladian operator. In the doubled Hilbert space this takes the form \cite{2020PhRvX..10b1019S,2015arXiv151008634A,Bhattacharya:2022gbz}
\begin{equation} \label{eq:ref}
\mathcal{L}_o=\left(I \otimes H_0-H_0^T \otimes I\right)+\frac{i\,\gamma}{2} \sum_i\left(I \otimes L_i^{\dagger} L_i+L_i^T L_i^* \otimes I-2 L_k^T \otimes L_i^{\dagger}\right),
\end{equation}
where $H_0$ is the Hamiltonian for the associated closed system Hamiltonian (which may also contain non-Gaussian terms) and all other components are as defined in Sec.~(\ref{sec2}). As mentioned there, we will primarily on those processes dictated by energy dephasing as in \cite{Cornelius_2022} so that the Lindbladian operator (\ref{eq:ref}) takes the form,
\begin{equation}
\mathcal{L}_o=\left(I \otimes H-H^T \otimes I\right)+\frac{i\gamma}{2N^2} \left(I \otimes L^{\dagger} L+L^T L^* \otimes I-2 L^T \otimes L^{\dagger}\right)\,,
\end{equation}
where, $L=L^{\dagger}=w(H)$ and we choose the power-law dependence $w(H_0)=H_0^{\delta}$ where $\delta$ can be either an integer or a half-integer \cite{Cornelius_2022}.\\ 

\begin{figure}[htb!]
    \centering
\includegraphics[scale=0.70]{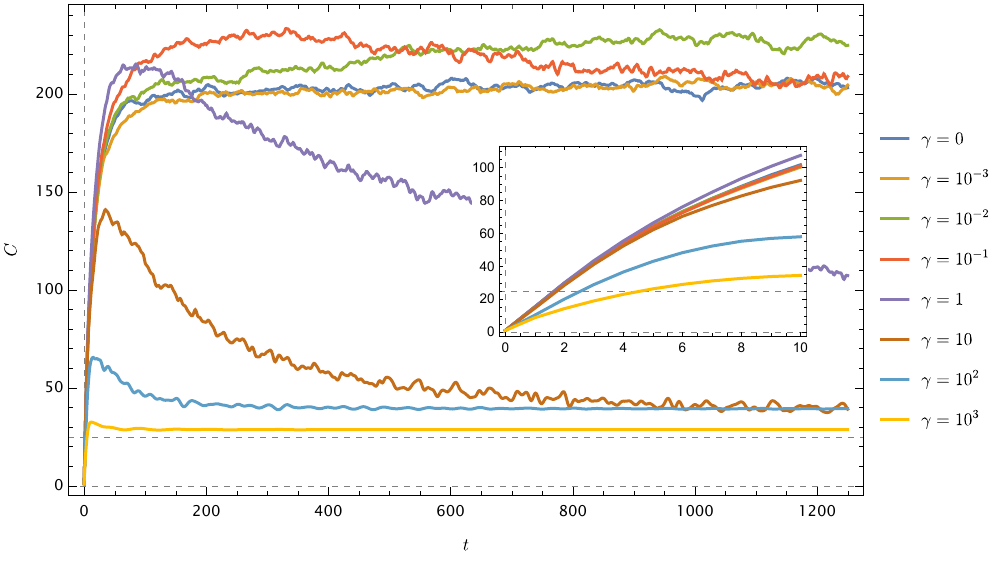}
     \caption{Spread Complexity for Gaussian RMT with noise.}  
         \label{fig-open}

\end{figure}

\noindent
Now we can compute the spread complexity following the procedure outlined in  Sec.~(\ref{sec4}) with one important caveat; unlike the Hamiltonian of the closed system, the Lindbladian $\mathcal{L} _0 $ is not Hermitian. Consequently, the orthogonalization procedure does not bring it into a tri-diagonal form; instead, the best we can do is to find a similarity transformation that puts it into the  \textit{upper-Hessenberg form},
\begin{equation} \label{hessenberg}
\mathcal{L}_h=\left(\begin{array}{cccccc}
a_0 & b_1 & c_1 & d_1 & \dots& \\
b_1 & a_1 & b_2 & c_2 & d_2\ \dots& \\
0& b_2 & a_2 & b_3 & c_3 &\dots \\
\ \ \ \ddots& \ \ \ \ddots &\ \ \ \ddots &\ \ \ \ddots & \ \ \ \ddots \\
\dots &  0& b_{N^2-2}  & a_{N^2-3} & b_{N^2-2} & c_{N^2-2}\\
\dots  & 0& 0& b_{N^2-2} & a_{N^2-2} & b_{N^2-1} \\
\dots&0 &0 &0 & b_{N^2-1} & a_{N^2-1}
\end{array}\right)\,.
\end{equation}
in the orthonormal Krylov basis \footnote{Remember that the size of the Lindbladian in the doubled Hilbert space is $N^2\,.$}. Then proceeding as before, we time-evolve the initial state, $|K_0\rangle = \ket{\psi(0)}$ in the doubled Hilbert space by the Lindbladian operator projected to the Krylov space $\mathcal{L}_h$ and, using (\ref{eq:complexity}), we can compute the spread complexity by computing the expectation value of $\widehat{K}$. Another consequence of the non-Hermicity of the Lindbladian is that we have to normalize the time-evolved state at each step of the time evolution. This computation has been done with $N^2 \times N^2$ square matrices with $N=25$, and with an ensemble averaging over $n=50$ realizations. Note that $N$ is the size of the matrices acting on the original Hilbert space. This becomes $N^2$ in the doubled Hilbert space. In Fig.~(\ref{fig-open}), we have demonstrated the time evolution of the spread complexity for the open system described by an RMT with Gaussian noise.

 
\noindent
As in the non-Gaussian case above, the complexity of the open system exhibits a characteristic pattern of initial sharp linear growth followed by saturation, as can be seen from Fig.~(\ref{fig-open}). There are, however, a number of notable differences in the behaviour of the Krylov complexity relative to the non-Gaussian random matrix models. Chief among these is the robustness of its  early-time linear growth against changes in the noise parameter $\gamma$. We have tested this over a range of $\gamma\in[0,10^{3}]$, finding significant deviation in the slope only for $\gamma\approx 100$.
A second key point to note is the nontrivial dependence of the peak and saturation values of the complexity on the amount of noise in the system, as parameterized by $\gamma$. Since the underlying Gaussian random matrix model is chaotic, we expect, primed by the results in \cite{Rabinovici:2023yex}, a high saturation plateau. For small but non-zero values of $\gamma$ the height of this plateau increases indicating that the system becomes more chaotic. However, there is a turnaround around at $\gamma = \gamma_{\mathrm{c}}\approx 10^{-2}$ when the plateau height starts to decrease.\\

\noindent
One way to understand this complexity suppression is that turning on noise $\gamma\neq0$ expands the operator space, allowing the system to explore a simpler set of operators. Consequently, the presence of the environment (represented by the noise) can also alter the magnitude of complexity and hence the chaos properties and induce decoherence in the system. This should be viewed in the light of previous studies   \cite{Bhattacharyya:2022rhm}, in which it was shown that Nielsen's circuit complexity saturates when the open system undergoes maximal decoherence.\\

\begin{figure}[t!]
    \centering
    \includegraphics[scale=0.84]{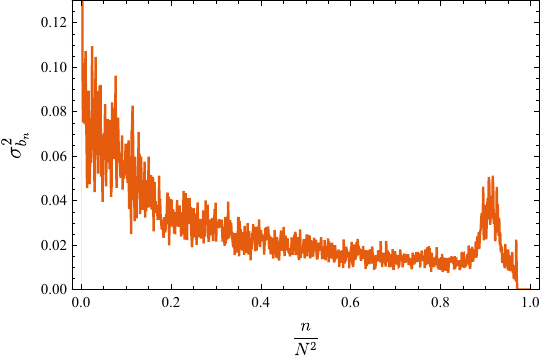}
     \includegraphics[scale=0.84]{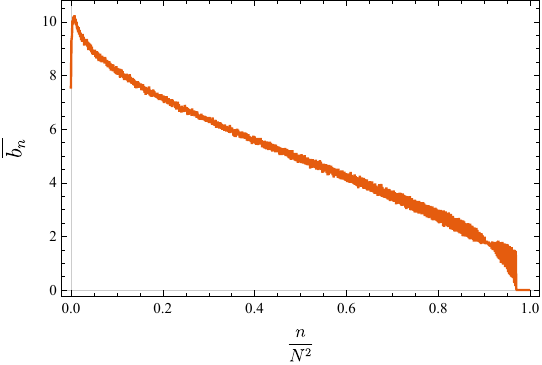}
    \caption{The mean $\overline{b_n}$ and the variance $\sigma_{b_n}$ of $b_n$ for Gaussian RMT with noise. We have set $\gamma=1.$ $N^2$ denotes the size of the Lindbladian.}
    \label{fig:sigmavariance}
\end{figure}

\noindent
As a final point, we investigate the distribution of elements of the Lindbladian $\mathcal{L}_h$ in upper-Hessenberg form for the Gaussian random matrix model with noise. We find that both the mean and variance vanish for the diagonal elements $a_n$ but are non-zero for $b_n.$ coefficients that fill in the first off-diagonal set. This is shown in Fig.~(\ref{fig:sigmavariance}). It is also interesting to note that, the variance of $b_n$ is typically non-zero in the presence of noise and vanishes in the $n \to N^2$ limit\footnote{This should be contrasted carefully with that of \cite{Balasubramanian:2022dnj} keeping in mind that we work in a doubled Hilbert space unlike \cite{Balasubramanian:2022dnj}.} We have checked this also for non-Gaussian RMT with noise as well as for various values of quartic and sextic couplings obtaining the same behaviour for the mean and the variance of $b_n$. Organizing our observations by three characteristic time-domains: the early-time behaviour, mid-time behaviour, and the saturation regime, we summarize the behaviours of the spectral form factor and spread complexity in Table.(\ref{tab:summary}) above.

\begin{table}[b!]
	\centering
	\setlength{\leftmargini}{0.3cm}
\begin{tabular}{|m{2.5cm}|m{6cm}|m{6cm}|}
	\hline \rowcolor{gray!30!} & { \textbf{Spectral Form Factor}} & { \textbf{Krylov Complexity}} \\
	\hline \textbf{   \begin{flushleft}
		{\small 	Quartic Non-Gaussian RMT}
	\end{flushleft} } & \begin{itemize}
	{\footnotesize 	\item Unaffected at early times.
		\item Increase in the size and onset of the dip with increasing coupling.
		\item Late-time plateau is unaffected by
		couplings.}
		\end{itemize} & \begin{itemize}
		{\footnotesize  \item Early-time growth varies with coupling.
		\item Peak complexity decreases wit increasing coupling.
		\item Late-time saturation is unaffected by  changing couplings.}
	\end{itemize} \\
	\hline  \textbf{{\small \begin{flushleft}
				Sextic Non-Gaussian RMT
	\end{flushleft}}} & \begin{itemize}
	{\footnotesize 	\item Early-time plateau decreases with increased coupling.
		\item Decrease in the depth of the dip with increasing coupling.
		\item Late-time plateau is unaffected by  couplings. }
	\end{itemize} & \begin{itemize} 
{\footnotesize 	\item Early-time slope increases with increasing coupling.
		\item Peak complexity decreases with increasing coupling.
		\item Late-time saturation increases with increasing coupling.}\end{itemize} \\
	\hline  \textbf{{\small \begin{flushleft}
				Gaussian RMT with Noise
	\end{flushleft}}}  & \begin{itemize}
	{\footnotesize 	\item Early-time behaviour is strongly dependent on noise.
		\item Dip height and time increases with increasing noise.
	\item Late-time plateau is unaffected by noise.} \end{itemize} & \begin{itemize} 
{\footnotesize 	\item Early-time growth weakly dependent noise. 
		\item Peak complexity depends non-trivially on noise.
		\item Late-time saturation values first increase then decrease with increasing noise}.\end{itemize} \\
	\hline
\end{tabular}
\caption{Contrasting the SFF and Spread Complexity for the various random matrix models.}
\label{tab:summary}
\end{table}

\section{Discussion}


Building on work initiated in \cite{Balasubramanian:2022dnj,Erdmenger:2023shk}, in this article we present a comprehensive comparative analysis of the spectral form factor (SFF) and spread complexity across various random matrix models. The SFF reflects the statistical properties of the energy spectrum and the level correlations, while the complexity measures the minimal spread of the wave function over all choices of basis. Our study serves to illustrate that the SFF and the complexity of spread of states (spread complexity) are indeed complementary measures of quantum chaos that capture different aspects of the quantum dynamics.\\

\noindent
Both quantities exhibit characteristic features, such as the slope-dip-ramp-plateau structure of the SFF and the linear-ramp-peak-saturation structure of the complexity, that depend on the non-Gaussianity and noise in the random matrix models in non-trivial ways. Since these features reveal the onset and saturation of quantum chaos, as well as the sensitivity to spectral rigidity and survival amplitude, our study emphasizes the significance of considering {\it both} the SFF and complexity in analyzing quantum chaotic systems, as they provide complementary information about the quantum dynamics and its relation to classical chaos.\\

\noindent
A key motivating question for our study was to clarify the effect of noise on quantum chaos. As is well-known, noise can affect quantum chaos in various ways, depending on the type and strength of the noise, the nature and size of the quantum system, and the observable of interest. For example, it was shown in \cite{casati_chirikov_1995} that in certain chaotic systems such as quantum kicked rotors, noise can induce decoherence and dissipation, which can destroy dynamical localization and restore classical diffusion. In other words, the presence of noise can make the quantum system behave more like its classical counterpart, which, in a sense can be interpreted as a decrease of quantum chaos. However, this does not mean that noise always has a negative effect on quantum chaos. In fact, there are also cases where noise can enhance or even induce quantum chaos, such as in noise-induced synchronization or noise-induced transitions between different types of synchronization . Therefore, the answer to this question may depend on the specific system, the type and strength of the noise, and the measure of quantum chaos one uses to probe the system.\\

\noindent
Our results here serve to further support this statement. Even in as simple a system as a Gaussian single random matrix model, depending on the strength of the noise and the measure used, the chaos properties of the quantum system are notably different (see Table(\ref{tab:summary})). Chief among these is the fact that the SFF, and by extension, the spectral properties of the Hamiltonian, appears to be blind to a possible transition in the chaoticity of the random matrix model from increasing to decreasing as the amount of noise is increased, as reflected in the saturation levels of the spread complexity (see Fig.(\ref{fig-open})). Beyond theoretical studies, noise plays a pivotal role in near future technologies such as quantum computing. For example, recent work by Wang {\it et.al.} \cite{Wang:2021iti} demonstrated how noise can be used to identify the most robust quantum circuit for a particular computing task, and to generate a mapping pattern tailored to the qubits of a targeted quantum device in a process they called Quantum Noise-Adaptive Search (QuantumNAS) with the potential to dramatically improve the accuracy of machine learning and quantum chemistry tasks on noisy quantum devices. As such, it is of immediate interested to sharpen our understanding of the interaction of such fundamental quantum systems such as random matrix models with noisy environments.\\

\noindent
There are a number of future directions worth pursuing. We will elaborate on a few of these below:
\begin{itemize}
    \item The focus of our study has been on just one random matrix ensemble; the GUE. It should be straightforward to extend our analysis to other ensembles. In particular, it will be very interesting to extend this study to non-Hermitian (Ginibre) random matrices. Such ensembles are complex-valued and related to the transfer matrices of patially-extended many-body quantum chaotic systems. This behavior is a consequence of the strongly interacting and spatially extended nature of the system, and is markedly different from the traditional emergence of Hermitian random matrix ensembles \cite{2020PhRvX..10b1019S,Shivam:2022fek}. More generally, such many-body systems exhibit complex and universal behaviors that challenge the fundamentals of our understanding of quantum dynamics and thermodynamics.
    \item Random matrix models are important because they serve as simple and tractable models of complex and realistic systems that are otherwise very difficult to analyze or even simulate. As simple as they are, these models are still rich enough to capture the essential features and behaviors of these systems, features such as spectral statistics, eigenvalue repulsion, phase transitions, and universality classes.  As such, another direction would be to investigate the physical implications of our findings, such as the relation between non-Gaussianity, noise, and quantum information scrambling or entanglement generation in quantum chaotic systems. Given the recent surge of interest in quantum information in the context of black holes and quantum gravity, we anticipate that this would likely be appreciated by a broad audience.
    \item Finally, our investigations in this paper are primarily numerical. Following \cite{Balasubramanian:2022dnj,Erdmenger:2023shk}, it will be of value to derive analytical results for the case of random matrix models with noise. In particular, it was shown in \cite{Erdmenger:2023shk} that spread complexity satisfies an Ehrenfest Theorem, with the SFF  driving its growth in a closed system. Do similar statements hold when noise is present in the system? It will be interesting to generalize the analytical arguments presented in \cite{Erdmenger:2023shk} to the Lindbladian system studied above.
\end{itemize}

\noindent
This article reflects our attempt to understand the subtle interplay between the spectral properties of random matrix models and quantum complexity, and what it teaches us about the underlying physical systems, particularly in the presence of noise. We believe that further exploration of this interplay between the SFF and complexity in diverse quantum systems will deepen our understanding of the underlying quantum dynamics and open up new avenues for the study of complex quantum phenomena. 

\section*{Acknowledgement}
We thank Pratik Nandy for initial collaboration on, and stimulating discussions related to this project; particularly for drawing to our attention the pivotal problem of noise in the RMT. We also thank Nitin Gupta and Jaco Van Zyl for many useful conversations on all things Krylov. A.B like to thank the FISPAC Research Group, Department of Physics, University of Murcia, especially, Jose J. Fernández-Melgarejo for hospitality and useful discussion based on the seminar given on Krylov complexity. A.B  is supported by Mathematical Research Impact Centric Support Grant (MTR/2021/000490) by the Department of Science and Technology Science and Engineering Research Board (India) and Relevant Research Project grant (202011BRE03RP06633-BRNS) by the Board Of Research In Nuclear Sciences (BRNS), Department of Atomic Energy (DAE), India. A.B thank the speakers and participants of the workshop ``Quantum Information in QFT and AdS/CFT-III" organized at IIT Hyderabad between 16-18th September, 2022 and funded by SERB through a Seminar Symposia (SSY) grant (SSY/2022/000446) and  ``Quantum Information Theory in Quantum Field Theory and Cosmology" between 4-9th June, 2023 hosted by Banff International Research Centre at Canada for useful discussions. JM would like to acknowledge support from the ICTP through the Associates Programme, from the Simons Foundation through grant number 284558FY19 and from the “Quantum Technologies for Sustainable Devlopment” grant from the National Institute for Theoretical and Computational Sciences of South Africa (NITHECS). G.J. is supported by the “Quantum Technologies for Sustainable Development" project from the National Institute for Theoretical and Computational Sciences
(NITHECS).We gratefully acknowledge support from the University of Cape
Town Vice Chancellor’s Future Leaders 2030 Awards programme which has generously funded this research and
support from the South African Research Chairs Initiative of the Department of Science and Technology and
the National Research Foundation.

\newpage 

\bibliographystyle{utcaps}


\bibliography{main}

\end{document}